\documentclass[twocolumn,showpacs,preprintnumbers,amsmath,amssymb]{revtex4}
\usepackage{graphicx}
\usepackage{dcolumn}
\begin{document}

\title
{Freezing of $^4$He and its liquid-solid 
interface from Density Functional Theory}

\author{F.~Ancilotto$^{1}$, M.~Barranco$^{2}$, F.~Caupin$^{3}$, R.~Mayol$^{2}$, and M.~Pi$^{2}$}
\affiliation{
$^{1}$ INFM-DEMOCRITOS 
and Dipartimento di Fisica ``G. Galilei'', Universit\`a di Padova, via Marzolo 8, I-35131 Padova, Italy\\       
$^{2}$ Departament ECM, Facultat de F\'{\i}sica, Universitat de Barcelona, E-08028, Spain\\
$^{3}$ Laboratoire de Physique Statistique de l'\'Ecole Normale Sup\'erieure\\
associ\'e aux Universit\'es Paris 6, Paris 7 et au CNRS, 24 rue Lhomond 75231 Paris Cedex 05, France} 
\date{\today}

\begin{abstract}
We show that, at high densities, 
{\it fully variational} solutions of solid-like type
can be obtained from a density functional formalism
originally designed for liquid $^4$He.
Motivated by this finding, we propose an extension of the method
that accurately describes the solid phase
and the freezing transition of liquid $^4$He at zero temperature.
The density profile of the interface between 
liquid  and the $(0001)$ surface of the $^4$He crystal is also
investigated, and its surface energy evaluated. The interfacial 
tension is found
to be in semiquantitative agreement with experiments 
and with other microscopic calculations.
This opens the possibility to use unbiased DF methods 
to study highly non-homogeneous systems,
like $^4$He interacting with strongly attractive 
impurities/substrates, or the nucleation 
of the solid phase in the metastable liquid.
\end{abstract}
\maketitle

\section{Introduction}

Helium crystals represent a fascinating system in which general properties of
crystalline surfaces such as the equilibrium crystal shape, surface phase transitions
(roughening) and elementary mechanisms of the crystal growth can be studied over a wide
temperature $(T)$ range, in principle down to absolute zero.
Helium crystals can be grown chemically and isotopically pure, and with very few lattice
defects. Helium crystal growth is fast in comparison to that of other solids
and it is characterized by the unique -quantum- phenomenon of crystallization waves,
i.e., melting-freezing waves which can easily propagate
on the liquid-solid interface at low temperatures. This allows an accurate measurement of the surface stiffness 
$\gamma$ and surface tension $\sigma$ and of their anisotropy~\cite{Andr91}. 
The research on the surface of helium crystals has been recently reviewed~\cite{Bali05}.

A microscopic approach to the study of solid $^4$He and 
liquid-solid coexistence at low temperature needs a fully quantum approach.
Accurate Path Integral Monte Carlo~\cite{draeger} (PIMC) 
in the 5-30 K 
temperature range
and Green Function Monte Carlo ~\cite{Kalo81} (GFMC) calculations
on solid $^4$He have been reported in the past.
More recently, the liquid-solid interface has been
adressed within a Variational Monte Carlo (VMC) approach
using shadow wavefunctions~\cite{Pede94}.

Density Functional (DF) methods~\cite{Evan79} represent
a useful computational tool to study the properties
of quantum inhomogeneous fluids, especially for
large systems where DF provides a good compromise
between accuracy and computational cost. 
Indeed, the $T=0$ properties of inhomogeneous
liquid $^{4}$He can be accurately described within  
DF theory (DFT) by using the phenomenological
functional proposed in Ref.~\cite{Dupo90} and later improved in
Ref.~\cite{Dalf95}, and similar ones. They have been widely used in a
variety of problems involving inhomogeneous $^{4}$He systems
like, e.g., liquid-vapor interface~\cite{Dupo90,Dalf95}, pure and doped clusters~\cite{Dalf95,Dalf94}, 
layering and prewetting transitions in films~\cite{Dalf95},
alkali atom adsorption on the surface of liquid $^4$He and droplets~\cite{Anci95}, 
vortices in $^4$He clusters~\cite{Dalf00}, etc.

In view of its conceptual interest on the one hand,
and of the  potential applications on the other hand,
we have undertaken a fully variational study of the
liquid-solid transition and coexistence of $^4$He at zero temperature within DFT.
Previous attempts on $^4$He~\cite{Dalf91,Moro91,Liko97,Caup04} or quantum
hard spheres~\cite{Dent90,Dent91} have been reported. They adapted the
techniques initially designed to address the liquid-solid transition in
classical systems (for a review see, e.g., Ref.~\cite{Oxto91}). They use trial
functions (sums of gaussians) to parametrize the solid density, and resort
to a second-order expansion of the energy around the liquid density
(Ramakrishnan-Yussouff (RY) approach)~\cite{Dalf91,Moro91,Liko97,Caup04},
or to a modified weighted density approximation exact to second order for
small perturbations of the liquid~\cite{Dent90,Dent91}. 
These methods need the density-density static linear response function
$\chi$ of the liquid as input. In
helium, these approaches have been found to be, at most, a reasonable
starting point to qualitatively describe the freezing
transition~\cite{Dalf91}; moreover, some of these attempts always give the
liquid~\cite{Moro91} or the solid~\cite{Liko97,Caup04} as the stable
phase.
Additional problems arise in treating the liquid-solid interface, 
because it is found to be too narrow to allow the use of
coarse graining methods~\cite{Caup04}. Besides, the use of trial function
density profiles makes all these approaches of limited applicability for
systems characterized by the presence of both solid and liquid phases.

\section{The Density Functional for solid $^4$He}

We attempt here a {\it fully variational} DF description of {\it both}
liquid and solid phases on the same footing.
Our starting point is a simplified version of the Orsay-Trento (OT)
functional~\cite{Dalf95}, where the energy of the $^4$He system 
is written as

\begin{eqnarray}
E[\rho ]=
&&\frac{\hbar^2}{2 m}
\int d{\bf r} \left( \nabla
\sqrt{\rho}\right)^2 +
\frac{1}{2} \int d{\bf r} \, d{\bf r}'\rho ({\bf r})
\rho ({\bf r}')V_\mathrm{LJ}(|{\bf r}-{\bf r}'|)\nonumber\\
+&&\frac{c_2}{2}\int d{\bf r}\rho ({\bf r})\bar{\rho}^2({\bf r} )
+\frac{c_3}{3}\int d{\bf r}\rho ({\bf r}) \bar{\rho}^3({\bf r})
\label{eq1}
\end{eqnarray}
The first term is the usual quantum kinetic energy term,
the second term contains a Lennard-Jones He-He pair
potential $V_\mathrm{LJ}(r)$ screened at distances shorter than a
characteristic length $h$. In the third and fourth terms,
the weighted density $\bar{\rho}$ is the average of the density
over a sphere of radius $\bar{h}$.
These terms account phenomenologically for short range correlations.
The parameters $h$, $c_2$ and $c_3$ 
(in the original formulation, $h=\bar{h}$)
are fixed to reproduce
the experimental density, energy per atom and compressibility
for the liquid at zero temperature and pressure~\cite{Dalf95}.
The original non-local kinetic energy term  proportional to
$\nabla \rho({\bf r}) \cdot \nabla' \rho({\bf r}')$~\cite{Dalf95}
has been dropped, because, as one might expect, it leads to 
unavoidable instabilities in the solid phase. Its neglecting
causes the static response function $\chi$ of the liquid to be 
only qualitatively described. We should mention the existence of
an alternative parametrization of this term
that still reproduces the experimental static function,
while being free from instabilities~\cite{Szyb05}. 
As the resulting DF stems from a somewhat
different strategy than current DF's do,
we have preferred to keep using an OT-like functional,
introducing the changes mentioned below.

The equilibrium density $\rho ({\bf r})$ is obtained
by minimizing $E[\rho ]$ with respect to density 
variations, subject to the constraint of a constant number of 
$^4$He atoms $N$. This is achieved by evolving in the imaginary
time a non-linear Schr\"odinger equation for the order parameter 
$\Psi ({\bf r})\equiv \sqrt{\rho ({\bf r})}$, 
where the Hamiltonian is given by 
$H=-\hbar ^2 \nabla^2 /(2 m)+ U[\rho ]$. The effective potential
$U$ is defined in terms of the variational derivative of the 
last three terms in the energy 
functional Eq. (\ref{eq1}) with respect to $\Psi$~\cite{Dalf95}.

\begin{figure}[ttt]
\centerline{\includegraphics*[clip,width=8.5cm,angle=0]{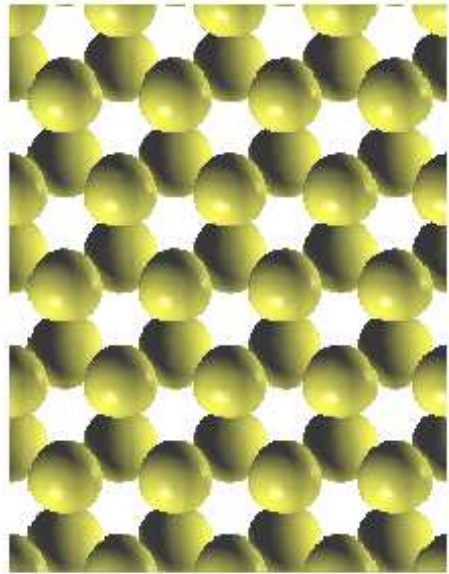}}
\caption{Surfaces of constant $^4$He density ($\rho =0.029\,$\AA$^{-3}$) of
a $^4$He hcp crystal, viewed along the c-axis.}
\label{fig1}
\end{figure}

The calculations are performed in a periodically
repeated supercell containing a fixed number of $^4$He atoms, and
the {\it starting} configuration
is a superposition of gaussian profiles with arbitrary but small
width, placed in the positions of the hcp structure, which is the
experimental solid structure of $^4$He. Interestingly,
for average densities corresponding to the stable liquid phase, 
the equilibrium density obtained is homogeneous, whereas for
higher densities -those roughly corresponding to the solid-
a solid $^4$He structure, characterized by a periodic density distribution with 
helium atoms at the lattice sites of an hcp crystal, is found to be the stable phase.
`Solid' means here a highly inhomogeneous configuration
characterized by regions, called `atoms' for brevity in the following,
where the density is very large. Such atoms are
only slightly overlapping with the density tails of neighboring atoms.
As an example, we show in Fig. \ref{fig1} the calculated density profile
for one such solid structure, computed at the experimental 
freezing density $\rho _f =0.0287$ \AA$^{-3}$.

We have studied other ordered structures as well.
An fcc lattice is also found to be a stable
phase at typical solid densities, 
its energy being only slightly higher than that of the hcp 
phase (by $0.02$ K per atom).
A simple cubic structure, at the same densities, 
is found instead to be unstable towards the 
homogeneous (liquid) phase.
We want to point out that the correct solid structure is found even if 
the initial gaussian superposition has not the correct 
parameters. For instance, we found that by starting with
gaussians, placed on a hcp lattice with a lattice constant 
twice the equilibrium one, but such that the total
normalization gives the correct $^4$He density, the equilibrium
hcp structure is eventually recovered at the end
of the minimization process. 

In spite of this encouraging result, a close analysis of the density
dependence of the total energy of the hcp solid $^4$He,
calculated using Eq. (\ref{eq1}), shows that it
largely deviates from the experimental result.
The reason is an unphysical, large density pile-up in the core
region of the $^4$He atoms, such that the resulting solid is too 
dense. To fix this unrealistic behavior,
we need a mechanism that makes energetically costly such
density pile-up. The simplest one is to add a `penalty' energy
term to the functional, which has the following form:

\begin{equation}
E_{p}[\rho ]=C\int \rho ({\bf r}) f[\rho ({\bf r})]\, d{\bf r}
\label{eq2}
\end{equation}
Here $f[\rho({\bf r})]$ is a `switch' function which becomes appreciably
different from zero only when the density is larger than
a predefined value $\rho_\mathrm{m}$. One such function is

\begin{figure}[ttt]
\centerline{\includegraphics*[clip,width=\columnwidth,angle=0]{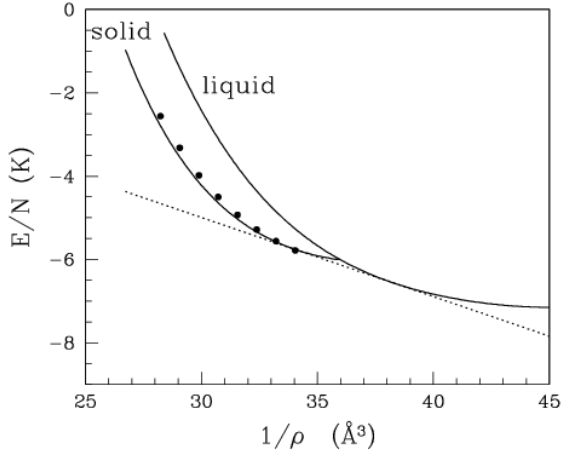}}
\caption{Solid lines: calculated solid and liquid EOS; dashed line: double tangent construction. 
Solid circles are the experimental data for the solid EOS~\cite{Edwa65}; the experimental EOS for the liquid
is not shown since it coincides, by construction, with that of the OT functional, whose agreement 
with experiment is excellent~\cite{Dalf95}.
}
\label{fig2}
\end{figure}

\begin{equation} 
f[\rho ({\bf r})]= 1+ \tanh \{ \beta [ \rho({\bf r})-\rho_\mathrm{m}] \} \; ,
\end{equation}
where $C$, $\beta $ and $\rho_\mathrm{m}$ are DF parameters.
The effect of $E_p[\rho]$ is to add 
to the effective potential $U[\rho]$ the term

\begin{equation} 
V_p({\bf r})=C \{ f[\rho ({\bf r})] + \beta \rho ({\bf r})
[1- \tanh^2 [ \beta (\rho ({\bf r}) - \rho_\mathrm{m} ) ] \} \; .
\end{equation}
If $C>0$ and when $\rho({\bf r}) \gtrsim \rho_\mathrm{m}$, this term acts as a
repulsive barrier, which forbids extra pile-up of the density. We can thus
use $C$, $\beta$, and $\rho_\mathrm{m}$ as adjustable parameters
in such a way to get agreement with the experimental 
equation of state (EOS)
for the solid phase. Note that there is
a large freedom in the choice of the 
penalty term since, by construction,
it has no effect whatsoever on the liquid 
structure, even in a highly structured liquid, 
because in practice $\rho_\mathrm{m}$ is 
a factor $\gtrsim 10$ than typical liquid densities.
We have found that the choice
$C=0.1$ Hartree, $\beta =40$ \AA$^3$,
and $\rho_\mathrm{m}=0.37$ \AA$^{-3}$~\cite{note1}, yields an EOS
for the solid in good agreement with experiments~\cite{Edwa65}, 
as shown in Fig.~2.

\section{The freezing transition and the liquid-solid interface}

Once the EOS of the homogeneous liquid and the EOS of the solid are determined, one can
easily locate the freezing transition by means of a double-tangent 
Maxwell construction.
We find a freezing pressure $P_\mathrm{f}=26.1$ bar
and a chemical potential at coexistence 
$\mu_\mathrm{f}=0.69$ K.
Our calculated values for $P_\mathrm{f}$ and the solid and liquid
densities at coexistence
are reported in Table \ref{table1}, together with those obtained by
other authors.

\begin{table}[ttt]
\caption{Freezing transition parameters ($\sigma=2.556$ \AA).\\
}
\label{table1}
\begin{tabular}{lccc}
Method & $\rho _l\sigma^3$ & $\rho _s\sigma^3$ & $ P_\mathrm{f}$ (bar) \\ \hline
Experiment~\cite{grilly}&   0.434 & 0.479 & 25.3 \\
GFMC~\cite{Kalo81}& 0.438 & 0.491 &  27.0 \\
VMC~\cite{Pede94}& 0.449 & 0.456 &  27.0 \\
RY~\cite{Dalf91}& 0.459 & 0.515  &  41.5 \\
RY~\cite{Caup04}& 0.435 & 0.513  & 25.7 \\
Present work & 0.437 & 0.490  & 26.1
\end{tabular}
\end{table}

We now turn to the liquid-solid interface of $^4$He, characterized by the interfacial tension 
$\sigma$ and width $\xi$. $^4$He is an example of a system 
for which $\sigma$ has been directly measured,
thanks to the possibility of propagating crystallization waves at
its surface. The value of $\sigma$ is $0.17\times 10^{-3}$ 
N/m~\cite{Andr91}, but $\xi$ is not known experimentally.

The $^4$He liquid-solid interface is an example of a system 
whose direct simulations by means of exact Monte Carlo methods
can be computationally very expensive.
Recently, it has been tackled by means of shadow wavefunctions VMC
\cite{Pede94}, and also by the rescaled DF we have referred to before 
\cite{Caup04}.
We have used the modified DF, Eqs. (\ref{eq1}-\ref{eq2}),
to study the liquid-solid interface.
To describe the coexisting phases, we start from a
slab of hcp crystal $^4$He in contact with a
region of liquid $^4$He, and
minimize the energy functional to find the equilibrium 
configuration,  imposing that during the 
minimization the chemical potential is constant and equal
to the value $\mu_\mathrm{f}=0.69$ K found from the double-tangent
construction illustrated in Fig. \ref{fig2}. 
We consider explicitly the interface between liquid $^4$He and 
the $(0001)$ surface of the crystal (basal plane).

\begin{figure}[ttt]
\centerline{\includegraphics*[clip,width=\columnwidth,angle=0]{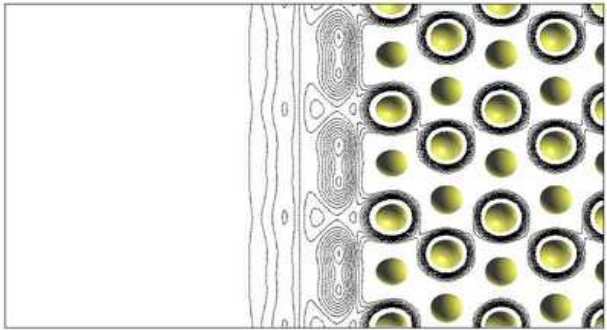}}
\caption{Liquid-solid interface shown by means of equal density contour lines  
(drawn between $\rho =0.02$ \AA $^{-3}$ and $\rho =0.05$ \AA $^{-3}$ )
in a plane perpendicular to the interface plane. Constant density surfaces 
(at $\rho =0.08$ \AA $^{-3}$)
are also shown to identify the atoms in the solid slab.}
\label{fig3}
\end{figure}

\begin{figure}[ttt]
\centerline{\includegraphics*[clip,width=\columnwidth,angle=0]{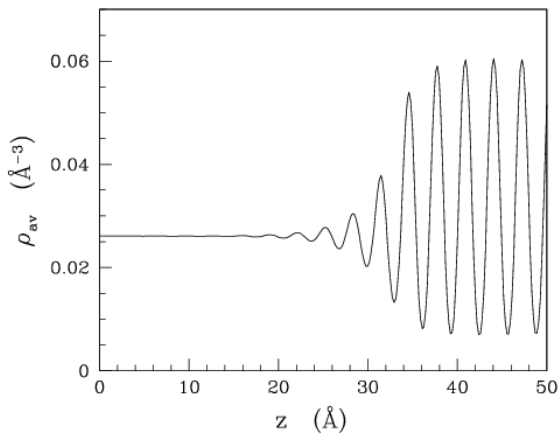}}
\caption{Planar-averaged $^4$He density plotted along the direction
normal to the interface. The solid is on the right, the liquid on the left.}
\label{fig4}
\end{figure}

Fig. \ref{fig3} shows a view of the equilibrium density
configuration at the liquid-solid interface.
A related quantity, the $^4$He density averaged over  
a plane parallel to the basal plane, is shown in Fig. \ref{fig4}.
The thickness $\xi$ of the liquid-solid interface has been
estimated by evaluating
the 10-90\% width of the curve interpolating the maxima in the 
density profile shown in Fig. \ref{fig4}.
We find $9.3\,$\AA, in close agreement with the VMC results.
Our findings are compared in Table II with other experimental and
theoretical results.

From the calculated density profiles we estimate
the liquid-solid surface tension as 
$\sigma=(E-\mu_\mathrm{f} N +P_\mathrm{f} V)/ (2A)$. Here
$V$ is the volume of the supercell, and $A$ is the surface area
of the exposed $(0001)$ face.
A factor 1/2 appears in the equation to account for
the two free surfaces delimiting the solid film in our slab geometry.
Our calculations, which are three dimensional in nature,
have been done by using a supercell accomodating $\sim 6$
layers of solid $^4$He and a thick liquid layer
in contact with it. We find that the convergence with respect to
the amount of liquid present in the supercell 
is rather slow 
(to obtain $\sigma$ accurately, one should
have two very wide liquid and solid regions in contact).
Our estimated 
value is $\sigma=0.1\times 10^{-3}$ N/m
\cite{note3}, which is a rather good
result, of quality similar to the VMC value~\cite{Pede94}.

\begin{table}[tbp]
\caption{Liquid-solid interface parameters.}
\label{table2}
\begin{tabular}{lll}
Method & $\sigma$ (10$^{-3}$ N/m) & $\xi$ (\AA )  \\ \hline
Experiment~\cite{Andr91}& 0.17 &  - \\
VMC ~\cite{Pede94}& 0.25$\pm$ 0.1 & 10-12.5 \\
DFT ~\cite{Caup04}& 0.47  & 5.6 \\
Present work & 0.1 &  9.3
\end{tabular}
\end{table}

\section{Conclusions}

We have shown that DF's currently used for liquid $^4$He admit
fully variational solutions of highly inhomogeneous type that
may represent localized helium `atoms'. With a simple modification,
one such DF is found to accurately describe both liquid and solid 
phases of $^4$He and their coexistence.
Our scheme does not
rely on any specific trial function density profile for the
solid phase, but gives an unbiased description of
both phases on the same footing.
This broadens the range of applicability of current DF,
permitting to study in an unbiased way 
highly non-homogeneous $^4$He systems, like 
droplets doped
with strongly attractive impurities, as for instance alkali 
ions \cite{Rossi04}, or $^4$He on strongly attractive substrates, 
such as graphite or carbon nanotubes \cite{graph03}.
Another potentially useful application is
the study of the nucleation of the
solid phase in the metastable superfluid~\cite{Maris03}.

We thank Franco Dalfovo for useful discussions.
F. A. aknowledges funding from CESCA-CEPBA, Barcelona,
through the program HPC-Europa Transnational Access.


\begin{thebibliography}{99}

\bibitem{Andr91} O. A. Andreeva and K. O. Keshishev, Physica Scripta, 
T{\bf 39}, 352 (1991).

\bibitem{Bali05} S. Balibar, H. Alles, and A. Y. Parshin, Rev. Mod.
Phys. {\bf 77}, 317 (2005). 

\bibitem{draeger}E.W.~Draeger and D.M.~Ceperley, Phys. Rev. B {\bf 61}, 12094 (2000).  

\bibitem{Kalo81} 
M. H. Kalos et al, Phys. Rev. B {\bf 24}, 115 (1981).

\bibitem{Pede94} F. Pederiva et al, 
Phys. Rev. Lett. {\bf 72}, 2589 (1994);
J. Low Temp. Phys. {\bf 101}, 543 (1995).

\bibitem{Evan79} R.~Evans, Adv. Phys. {\bf 28}, 144 (1979).

\bibitem{Dupo90} J.~Dupont-Roc et al, J. Low Temp. Phys. {\bf 81}, 31 (1990).

\bibitem{Dalf95} F.~Dalfovo et al, Phys. Rev. B {\bf 52}, 1193 (1995).

\bibitem{Dalf94} F. Dalfovo, Z. Phys. D {\bf 29}, 61 (1994);
M. Barranco and E. S. Hernandez,
Phys. Rev. B  {\bf 49}, 12078, (1994);
M. Casas et al, Z. Phys. D {\bf 35}, 67 (1995).

\bibitem{Anci95} F. Ancilotto et al, Z. Phys. B {\bf 98}, 323 (1995);
F. Stienkemeier et al, Phys. Rev. B {\bf 70}, 214509 (2004).

\bibitem{Dalf00} F. Dalfovo et al, Phys. Rev. Lett. 
{\bf 85}, 1028 (2000);
F. Ancilotto, M. Barranco, and M. Pi, {\it ibid.}
{\bf 91}, 105302 (2003).

\bibitem{Dalf91} F. Dalfovo et al, 
Europhys. Lett. {\bf 16}, 205 (1991).

\bibitem{Oxto91} D.W. Oxtoby, in \textit{Liquids, Freezing and Glass
Transition, Les Houches}, edited by J.P. Hansen, D. Levesque and J.
Zinn-Justin (Elsevier Science Publishers, 1991), p. 145.

\bibitem{Moro91} S. Moroni and G. Senatore, 
Europhys. Lett. {\bf 16}, 373 (1991).

\bibitem{Liko97} C. N. Likos, S. Moroni, and S. Senatore, Phys. Rev. B {\bf 55}, 8867 (1997).

\bibitem{Caup04} F. Caupin and T. Minoguchi, J. Low Temp. Phys. 
{\bf 134}, 181 (2004); {\it ibid.} {\bf 138}, 331 (2005).

\bibitem{Dent90} A. R. Denton et al., 
Phys. Rev. Lett. {\bf 64}, 1529 (1990); J. Phys. Cond. Mat. {\bf 3} 593 (1991).

\bibitem{Dent91} A. R. Denton, P. Nielaba, and N. W. Ashcroft, J. Phys. Cond. Mat. 
{\bf 9} 4061 (1997).

\bibitem{Szyb05} L. Szybisz and I. Urrutia, Phys. Lett A {\bf 338}, 155
(2005).

\bibitem{note1} We have also employed an $\bar{h}$ to compute the 
coarse-grained density in the DF Eq. (\ref{eq1})  that
is 6.5\% larger than that of Ref.~\cite{Dalf95}, while keeping
for the screening distance $h$ its original value.

\bibitem{Edwa65} D. O. Edwards and R. C. Pandorf, Phys. Rev. A 
{\bf 140}, 816 (1965).

\bibitem{grilly} E. R. Grilly, \textit{J. Low Temp. Phys.}, \textbf{11}, 33 
(1973).

\bibitem{note3} A similar value 
is obtained with the equivalent definition for $\sigma$ wich uses 
the difference between the actual interface system and a 
reference state, composed of two separated volumes 
of homogeneous solid and liquid, the volume of each being 
defined by the Gibbs dividing surface. 

\bibitem{Rossi04} M.Rossi et al., Phys. Rev. B {\bf 69}, 212510 (2004).

\bibitem{graph03} T.Wilson and O.E. Vilches, Low Temp. Phys. {\bf 29}, 732 (2003).

\bibitem{Maris03} H.J.~Maris and F.~Caupin, J. Low. Temp. Phys. {\bf 131}, 145 (2003).


 



\end{thebibliography}
\end{document}